\documentstyle[12pt,epsf,epsfig]{aipproc}

\begin{document}
\sloppy         
\title{\LARGE\bf Fiber technology applications for a future e$^+$e$^-$
linear collider detector}
\author{H. Leich, R. Nahnhauer, R. Shanidze\footnote{on leave from
 High Energy Physics Institute, Tbilisi State University}}
\address{DESY-Zeuthen, 15738 Zeuthen, Germany}

\maketitle
\begin{abstract}
      The advantages and possibilities of fiber technology for the
      detection of particles in 500 GeV e$^+$e$^-$ reactions are considered.       
      It is suggested to build a fast trigger which could be used also
      for intermediate tracking. A fiber preshower in front of the
      electromagnetic calorimeter would allow to identify electrons and
      photons with a space precision better than 100 $\mu$m.
\end{abstract}

\section*{Introduction}
e$^+$e$^-$ reactions have particular advantages for the study of particle
properties and interaction pecularities. That is the reason for the
worldwide discussion about problems and possibilities of future e$^+$e$^-$
linear colliders in the energy range between 500 -- 1500 GeV \cite{lit1}.
 
Detectors at such colliders have to handle very short bunch crossing
times and/or tremendous background rates originating from new sources
due to the strongly collimated beams. All these questions were studied
in detail during an Joint ECFA/DESY workshop in 1996 \cite{lit2}. The following
considerations arised from discussions during this workshop.

\section*{Reactions studied}
To investigate the capabilities of special detector configurations
we simulated benchmark e$^+$e$^-$ reactions at 500 GeV :

\begin{center}
\begin{tabular}{l@{\,\qquad $\rightarrow$\qquad}l@{$\quad$}c}
e$^+$e$^-$& $W^+W^-$& (1)\\
e$^+$e$^-$& t\={t}& (2)\\
e$^+$e$^-$& HX & (3)\\
\multicolumn{2}{l}{beamstrahlung} &(4)
\end{tabular}
\end{center}

\noindent
and traced the resulting secondary particles through the detector,
using the program systems PYTHIA \cite{lit3}, ABEL \cite{lit4} and GEANT \cite{lit3}.

The number of particles to be detected varies between 2 (pure leptonic
$W^+W^-$ decays) and 200. Averages are given in table 1. 

\begin{center}
\begin{table}[h!]
\caption{Average number of all particles, charged particles and leptons for W, top and higgs
             production. The event rates correspond to a luminosity
             L= 2 $\cdot$ 10$^{33}$ cm$^{-2}$ sec$^{-1}$}
\renewcommand{\arraystretch}{1.5}
\begin{tabular}{|c|c|c|c|c|} \hline 
&  $<$ N $>$ & $<$ N$_{ch}$ $>$ &  $<$ N$_{lep} >$  & ev. rate \\ \hline
 $W^+W^-$ & 57  &    26   &    0.9     &   1/min \\ \hline
 tt& 115 &    55   &    2.1    &  1/10min\\ \hline
 HX&  78 &    36   &    1.5    &  $\sim$1/h \\
\hline
\end{tabular}
\end{table}
\end{center}
\vspace{-0.8cm}
The kinematics is  different for all processes considered (see fig.
\ref{bild1}). $W^+W^-$ pairs are produced strongly in forward and backward
direction in contrast to top and higgs production. Particles
from beamstrahlung are mostly low energetic and bounded to the inner part
of the detector near to the beam pipe, if a magnetic field of 3 T is
assumed. 
    
\section*{Fast inner trigger}
\begin{figure}[t] 
\begin{minipage}[1]{14cm}
\begin{center}
\epsfig{file=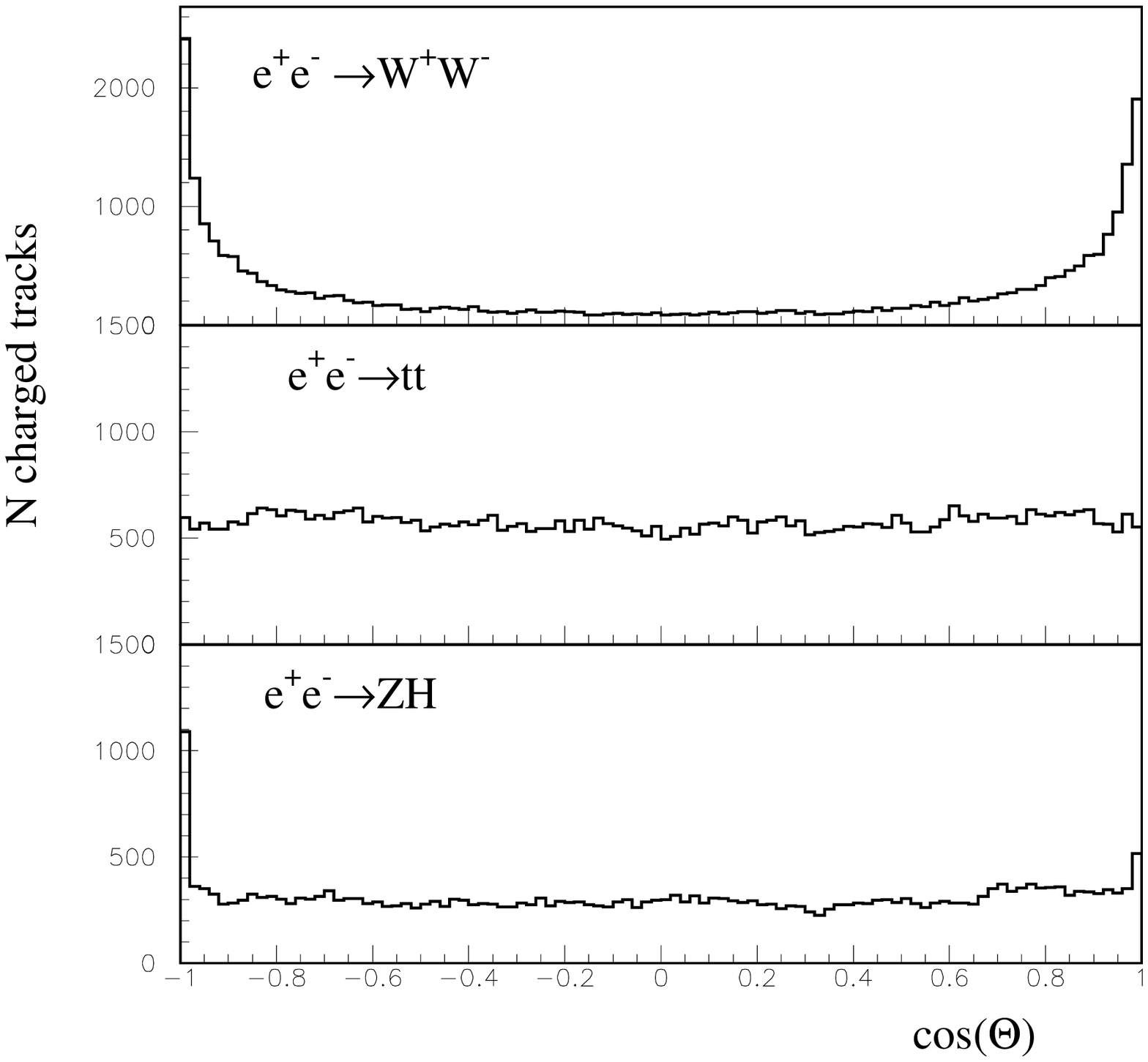,width=12cm,height=10cm}
\caption{ Angular distribution of secondary particles from $W^+W^-$,
  t$\bar t$ and higgs production reactions in 500 GeV $e^+e^-$ scattering}
\label{bild1}
\end{center}
\end{minipage}
\hfill
\vspace{1.5cm}
\begin{minipage}[b]{6cm}
\begin{center}
\epsfig{file=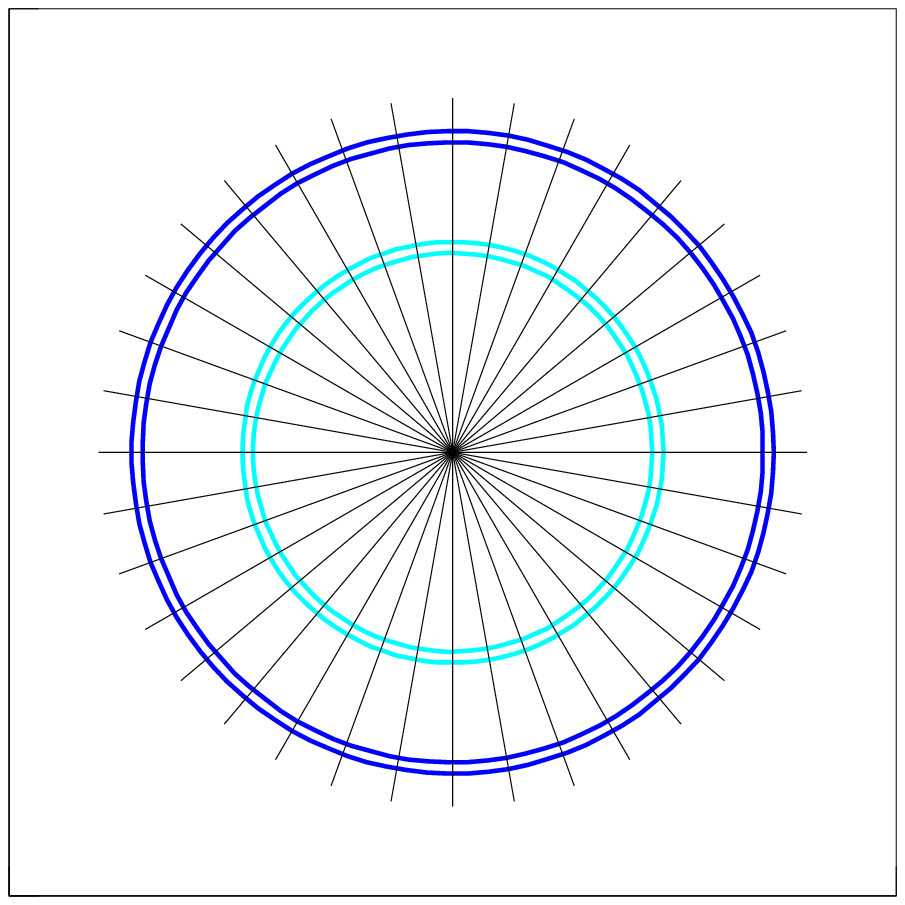,height=5.5cm,width=5.5cm}
\caption{Schematic r$\phi$-view of a possible central trigger structure}
\label{bild2}
\end{center}
\end{minipage}
%
%%\hspace{1.0cm}
\begin{minipage}[b]{8cm}
\begin{center}
\epsfig{file=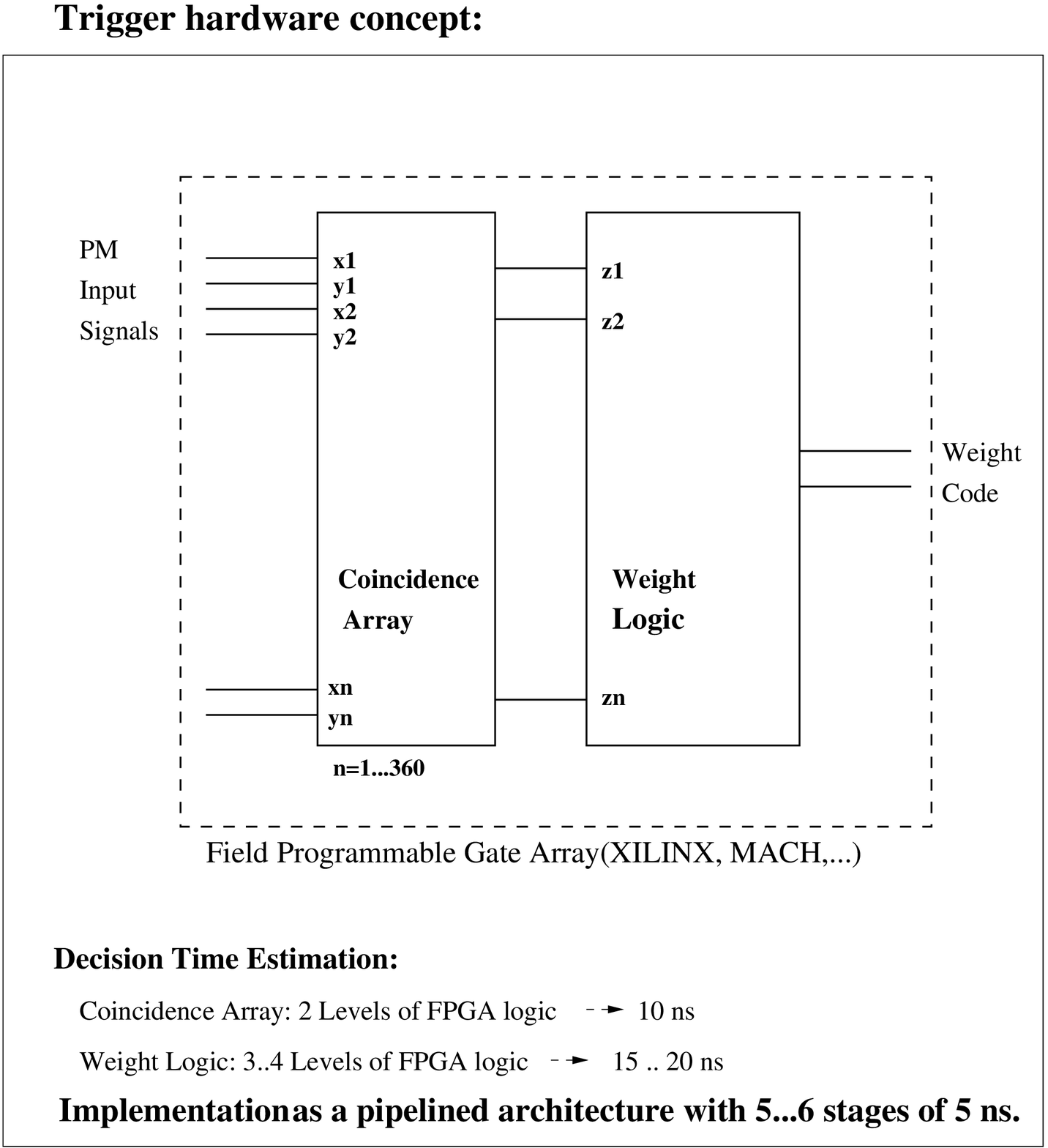,height=7.5cm,width=7.7cm}
\caption{Scetch of a trigger electronic scheme}
\label{bild3}
\end{center}
\end{minipage}
\hfill
\end{figure}

To suppress background and select events with high
transverse momentum jets, a fast inner trigger could be useful in
particular for collider operations with very short bunch crossing
times. We propose a layout as schematically drawn in fig. \ref{bild2}.

In a magnetic field of 3 Tesla two cylinders of 1 m length surround
the interaction point at r = 18 cm and r = 28 cm. Each cylinder consists of
4 layers of scintillating fibers of 1 mm diameter parallel to the
z--axis. The fibers are combined to form $\Phi$--slices for the inner and
outer cylinder. For a one degree resolution one would have to handle
therefore 2 $\cdot$ 360 channels. The light is read out by light guides coupled
to normal photomultipliers or by hybrid devices withstanding magnetic
fields \cite{lit5}. 

\begin{figure}[b] % fig 4,5,6
\begin{minipage}[t]{7cm}
\begin{center}
\epsfig{file=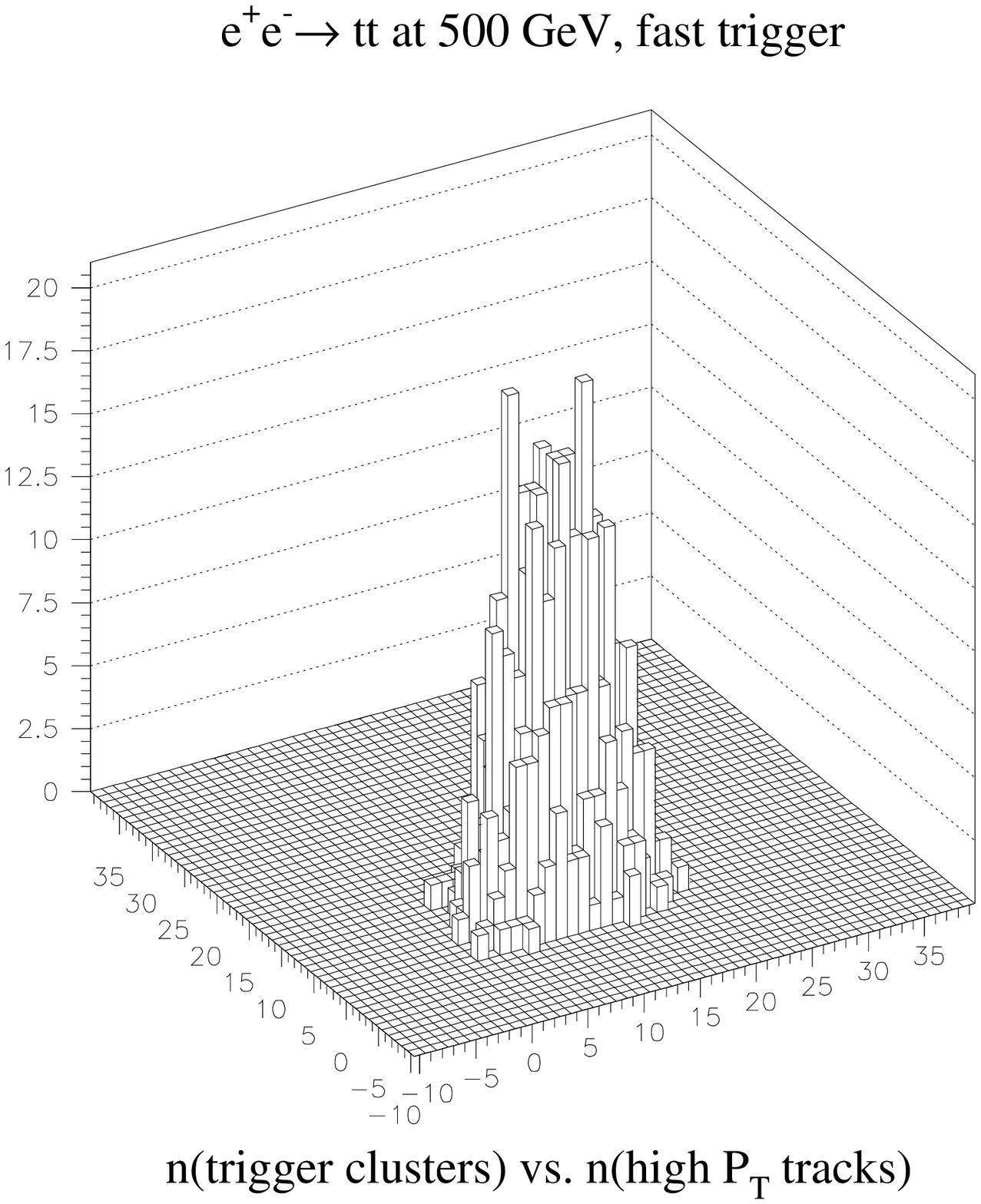,width=7cm,}
\end{center}
\caption{Number of high $P_t$ tracks vs. number of detected trigger
            clusters for 500 GeV t$\bar{t}$-production }
\label{bild4}
\end{minipage}
\hfill
\begin{minipage}[t]{7cm}
\begin{center}
\epsfig{file=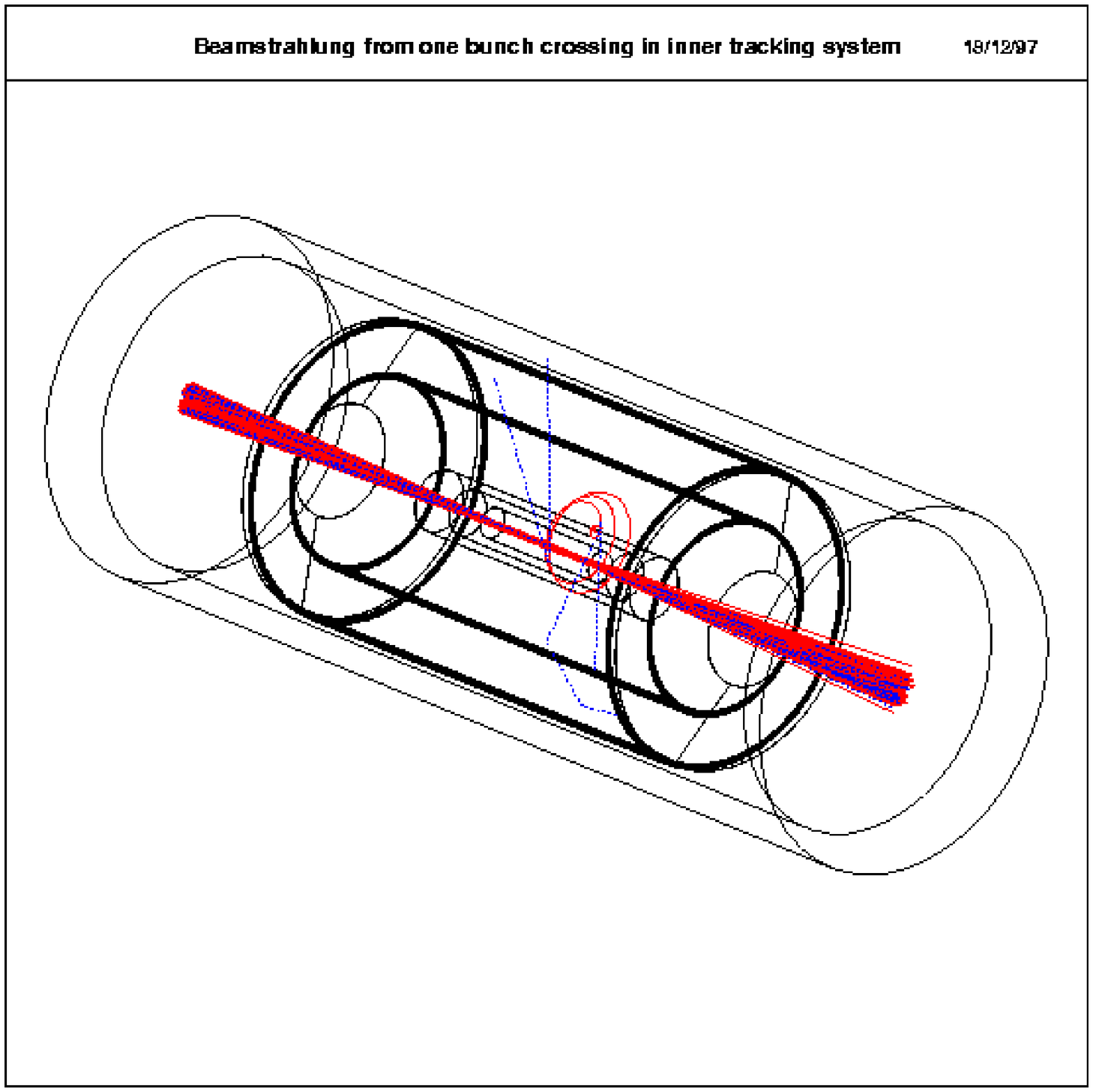,width=7cm}
\vspace*{0.0cm}
\caption{  Distribution of tracks from beamstrahlung produced
  in one bunch crossing in the inner tacking system in a 3T magnetic field}
\label{bild5}
\end{center}
\end{minipage}
\hfill
\begin{center}
\begin{minipage}[b]{12cm}
\epsfig{file=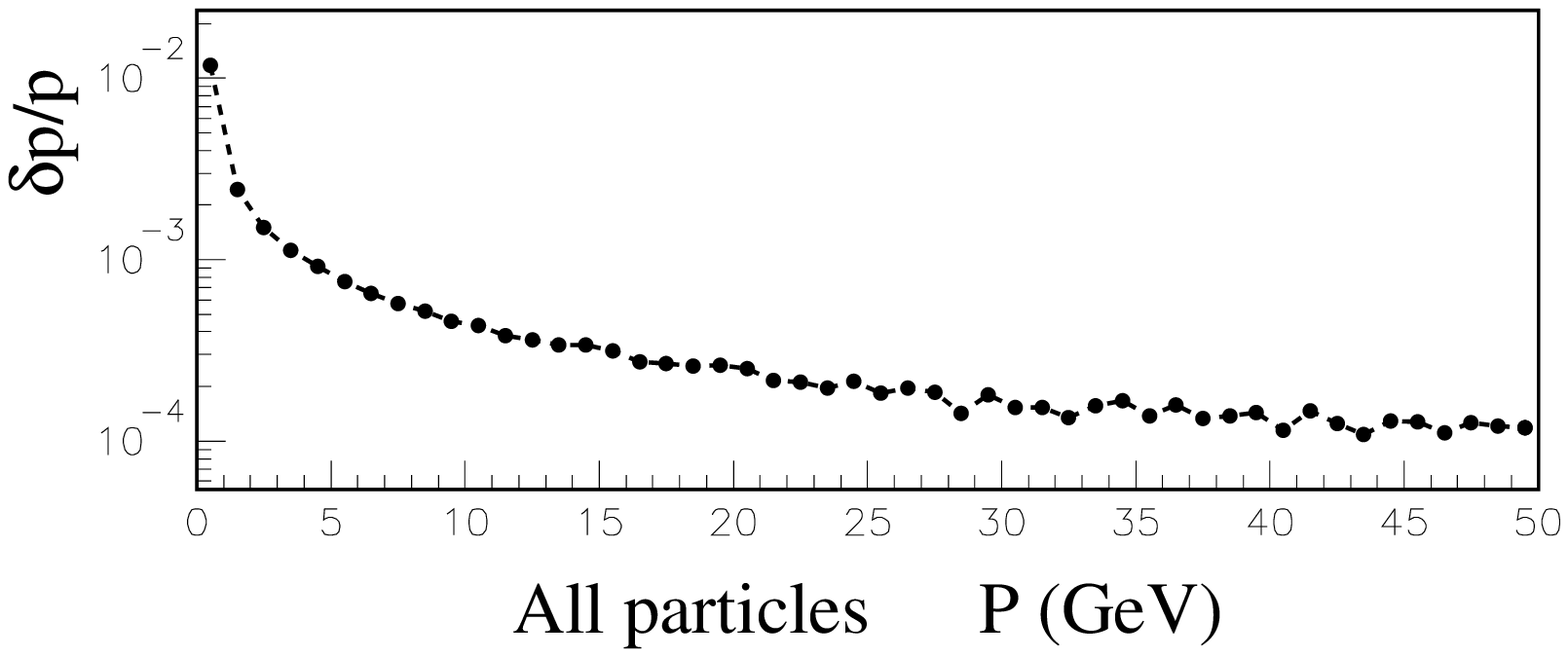,width=12cm}
\caption{ Relative precision of particle momenta after having passed
            the central inner fiber tracker }
\label{bild6}
\end{minipage}
\end{center}
\end{figure}

The simplest trigger condition is to demand signals above a certain
threshold from the same $\Phi$--segment of both cylinders. Trigger times
below 10 nsec seem to be in reach with the above arrangement. A
possible hardware scheme is given in fig. \ref{bild3}. It would allow even to
form clusters between neighboured slice signals.
  The trigger efficiency is limited only by geometry. For t\={t} and Higgs
production it is nearly 100 \%. The number of fake triggers may be
decreased by smaller cell sizes. With the configuration described, the
average occupancy is  5 \%($W^+W^-$), 14 \%(t\={t}) and 7 \%(HX). In
 fig. \ref{bild4} the clear
correlation between high p$_t$ tracks and trigger clusters observed is
demonstrated. No trigger is found using simulated tracks from
beamstrahlung. A typical particle distribution for this process in one
bunch crossing is shown in the r$\Phi$--plane in fig. \ref{bild5}.

\section*{Central inner tracker}
It is now easy to extend the mechanical trigger layout to build a
central fiber tracker. In addition to the fibers parallel to the z-axis
four layers of staggered fibers inclined by 11 degrees will be added
to both cylinders. With a pitch of 1.08 mm between adjacent fibers one
would have 1 050 and 1 650 fibers per layer for the inner and outer
cylinder respectively, adding up to a total number of 21 600 channels to
be read out e.g. via light guides and multianode photomultipliers \cite{lit6}
or VLPC's \cite{lit7}.

The space point resolution of such a configuration would
be $\Delta$ r$\Phi$ = 80 $\mu$m and $\Delta$z = 410 $\mu$m. The geometrical 
acceptance and average occupancy per layer is given in tab.2.

\begin{center}
\begin{table}[h!]
\caption{Geometrical acceptance and average occupancy for
             reactions 1-3 at the inner and outer cylinder of a
             central inner fiber tracker.}
\renewcommand{\arraystretch}{1.5}
\begin{tabular}{|c|c|c|c|c|c|c|} \hline 
 cylinder  & \multicolumn{3}{c|}{\qquad
  geometr. acceptance \qquad}&\multicolumn{3}{c|}{\qquad aver. 
occupancy/layer \qquad}\\ \hline 
 &  $W^+W^-$  &   t\={t}   &  HX   &
 $W^+W^-$  &   t\={t}   &  HX  \\ \hline
  inner   & 0.67  &  0.93  &  0.92 & 0.021 & 0.060  & 0.031 \\ \hline
  outer  & 0.51  &  0.86  &  0.84 & 0.011  & 0.038  & 0.019 \\
\end{tabular}
\end{table}
\end{center}
\vspace{-0.8cm}
Due to multiple scattering and energy loss in the fiber material the
described detector will naturally influence the precission in reach
for a following outer tracker. That seems to be important only for
particles with energies below 2 GeV, as can be seen from fig. \ref{bild6}.

\section*{ Fiber preshower}
A fiber tracker and preshower was first successfully used in the
UA2--experiment \cite{lit8}. We will closely follow that concept using however
only fibers in z--direction because no precise tracking in two
coordinates is necessary.

We suggest to build a fiber--lead sandwich cylinder with a radius of
1m and a total length of 7 m splitted in two parts in the middle. Using
fibers of 1mm diameter and 3.5 m length arranged with a pitch of 1.05 mm
this results in 6 000 fibers per layer and cylinder half. Six staggered
inner layers parallel to z would allow a resolution of 
$\Delta$ r$\Phi$ = 50 $\mu$m. 
After 9 mm of lead corresponding to 1.6 radiation lengths, four
staggered outer layers still would give $\Delta$ r$\Phi$ = 80 $\mu$m. 

\begin{figure}[b] % fig 7 and fig 8 and fig 9
\begin{minipage}[t]{7cm}
\begin{center}
\epsfig{file=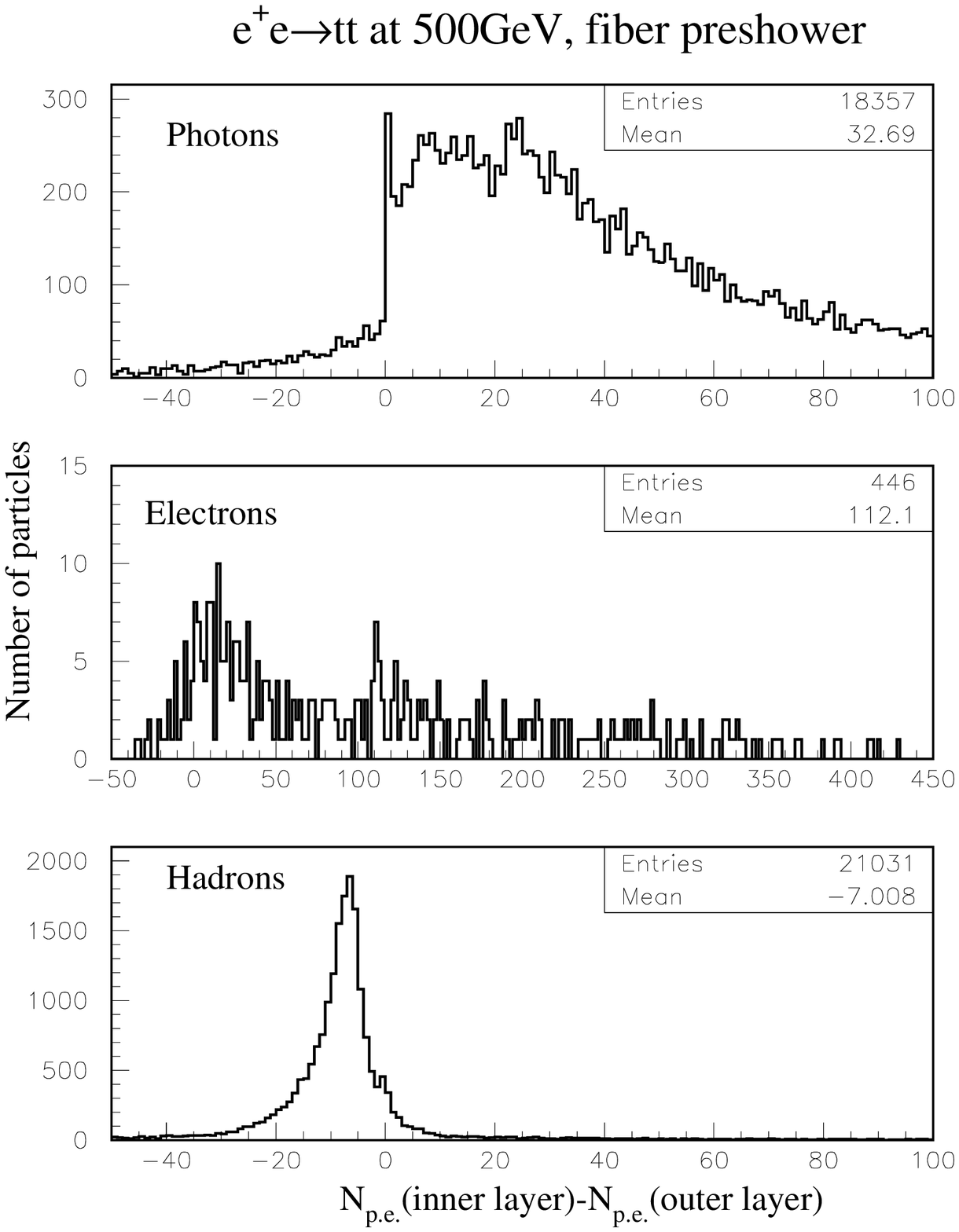,width=7cm,}
\end{center}
\caption{Difference of number of photoelectrons for inner and outer
         preshower layers normalized to equal numbers of them for
         photons, electrons and pions}
\label{bild7}
\end{minipage}
\hfill
\begin{minipage}[t]{7cm}
\epsfig{file=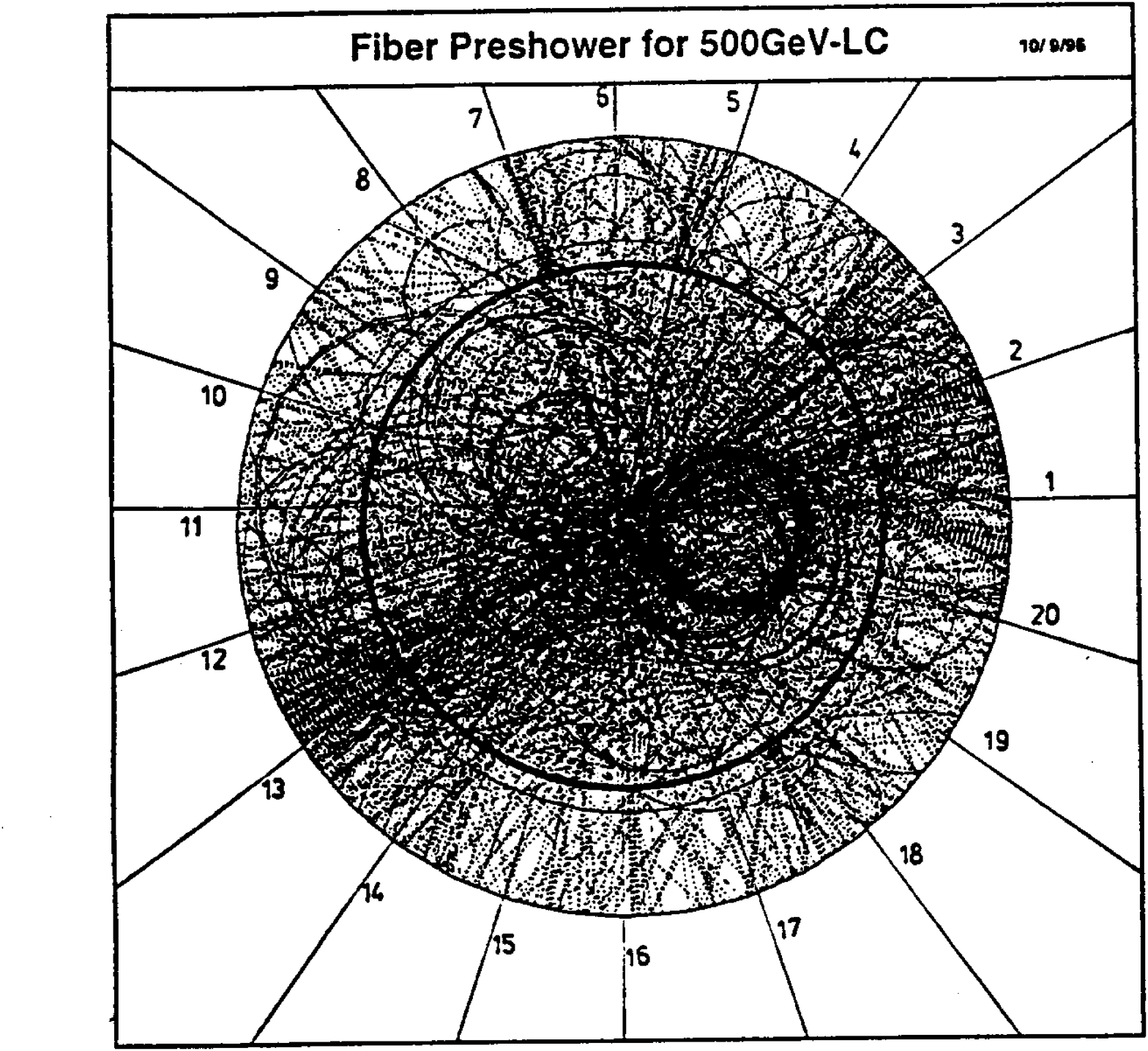,width=7cm}
\vspace*{1cm}
\caption[]{Typical event from t$\bar{t}$ production at 500 GeV as seen
  by the preshower}
\label{bild8}
\end{minipage}
\hfill
\vspace*{1.0cm}
\begin{center}
\begin{minipage}[b]{12cm}
\epsfig{file=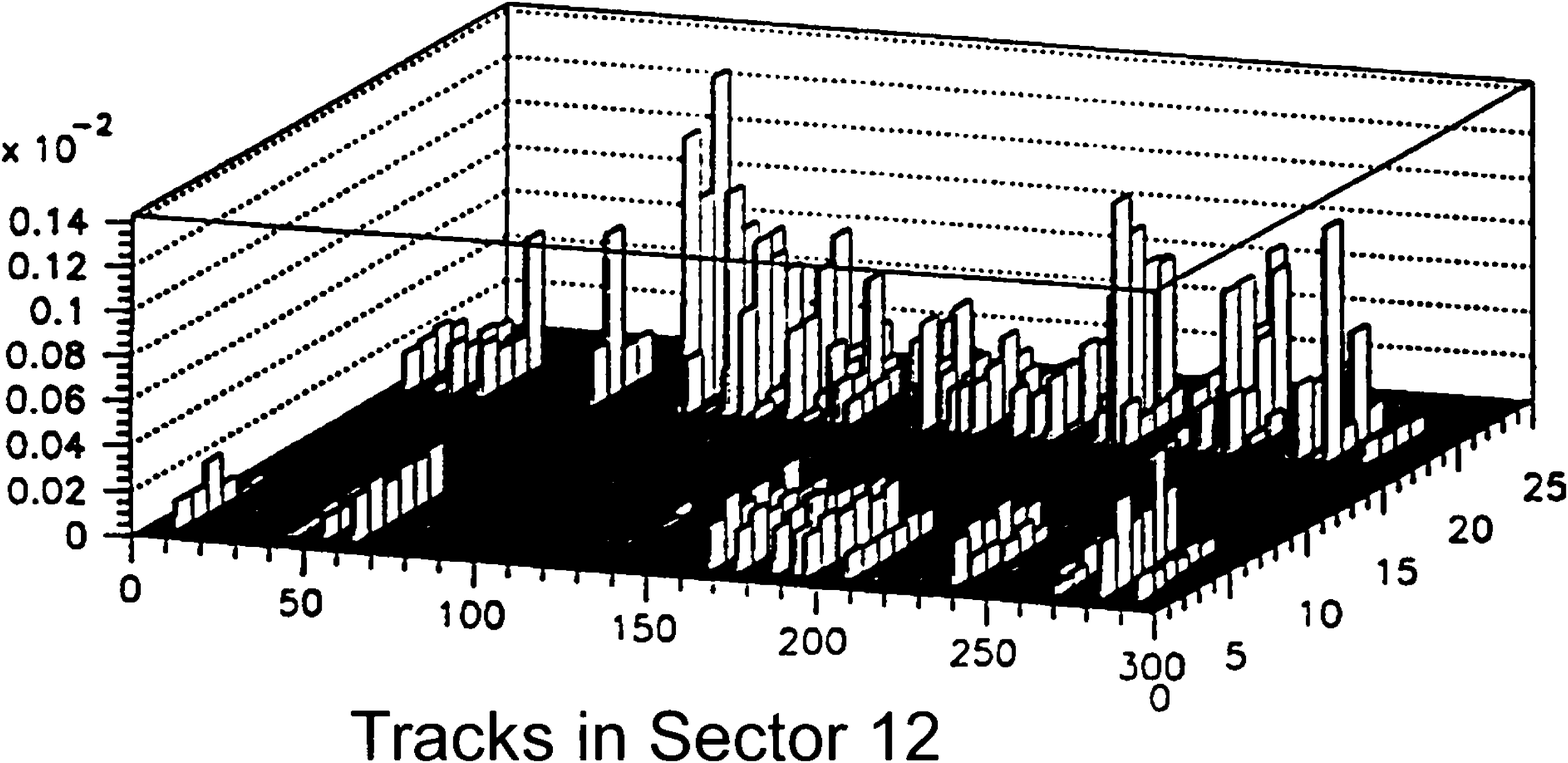,width=10cm}
\vspace*{1.0cm}
\caption{Particle tracks from preshower fibers weighted with the
            corresponding light amplitudes for sector 12 of fig. 8}
\label{bild9}
\end{minipage}
\end{center}

\end{figure}

The very good two track resolution and electron identification of
such a device allows excellent $\gamma$/e/$\pi$ separation important for the
precise measurement of many physical variables. Adding up the signals
of hitted fibers for the inner and outer preshower layers the
corresponding difference of the number of detected photoelectrons is
shown in fig \ref{bild7} for photons, electrons and pions. Correcting for
the different number of layers one finds integral values of
$\Delta$N$_{pe}$ of $\gamma$/e/$\pi$ = 33/105/0.

A typical t\={t}--production event as seen by the preshower is shown
in fig. \ref{bild8}.
In fig. \ref{bild9} the position of tracks weighted with  light
amplitudes from the hitted fibers is plotted for sector 12 of fig \ref{bild8}. One clearly can see the resolving power of the detector.

An open question is how to read out the 120000 channels of the
preshower. Keeping in mind the low rate of interesting events there is
no need to do it fast. Therefore image intensifier chains and CCD
already applied for UA2 and now used for the 10$^6$ channel tracker of
the CHORUS experiment \cite{lit9} may be a todays solution also here.

\section*{Summary}
Fiber technology provides various interesting detector applications
for a 500 GeV e$^+$e$^-$ collider.

A very fast inner trigger could select e.g. top and higgs
production events with high efficiency within about 10 nsec. Extended
to a two coordinate measuring inner tracker a space point resolution of
$\Delta$ r$\Phi$ = 80 $\mu$m and $\Delta$z = 410 $\mu$m could 
be reached. A nearly complete suppression of background of 
beamstrahlung seems to be possible.

 A one coordinate high resolution fiber preshower provides excellent
 $\gamma$/e/$\pi$ separation within $\Delta$ r$\Phi$ below 100 $\mu$m.

%%%%%%%%%%%%%%%%%%%%%%%%%%%%%%%%%%%%%%%%%%%%%%%%%%%%%%%%%
%       3 Pages of figures
%%%%%%%%%%%%%%%%%%%%%%%%%%%%%%%%%%%%%%%%%%%%%%%%%%%%%%%%%
%  Figs 1,2,3
%%%%%%%%%%%%%%%%%
%%% \clearpage
%%%%%%%%%%%%%%%%%%%%%%%%%%%%%%%%%%% New PAGE %%%%%%%%%%%%%%%%%%%%
%  Figs 4,5,6
%%%%%%%%%%%%%%%%%
\vspace*{-2cm}
%%%%%%%%%%%%%%%%%%%%%%%%%%%%% new page
%   Fig.7,8,9
%%%%%%%%%%%%%%%%%%  
%%%%\newpage
\vspace*{-2cm} 

\begin{references}
\bibitem{lit1}Loew, G.A., (ed.), International Collider Technical
Review Committee Report, SLAC-R-95-471, 1995
\bibitem{lit2}Brinkmann, R., et al.,(eds.), Conceptual Design of a
              500 GeV e$^+$e$^-$ Linear Collider with Integrated X-ray
              Laser Facility, DESY 1997-048/ECFA 1997-182
\bibitem{lit3}CERN Program Library Long Writeups W5035 and W5013
\bibitem{lit4}Yokoya, K., KEK report 85-9 (1985), see also:
              Schulte, D., preprint DESY-TESLA 97-08 (1997)
\bibitem{lit5}Arisaka, K., contribution to this workshop
\bibitem{lit6}B\"ahr, J., et al., Proceedings of the workshop SCIFI93,
      eds. A. Bross, R. Ruchty, M. Wayne, Notre Dame, USA, 1993,
      p. 183
\bibitem{lit7}Adams, D., et al., {\it Nucl. Phys.} {\bf B}
              (Proc.Suppl.) 44, 340, (1997)

\bibitem{lit8}Ansorge, R., et al., {\it NIM} {\bf 265}, 33, (1988)
\bibitem{lit9}Annies, P., et al., {\it NIM} {\bf A367}, 367, (1995)
\end{references}
\end{document}